\documentclass[journal]{IEEEtran}
\usepackage{graphicx}
\usepackage{float}
\graphicspath{{./Figures/}}
\usepackage[T1]{fontenc}      
\usepackage{booktabs}       
\usepackage{array}          
\usepackage{hyperref}
\ifCLASSINFOpdf
\else
\fi

\begin{document}
%
\title{Signal and Noise Classification in Bio-Signals via  unsupervised Machine Learning }
%
%
%

\author{Sansrit Paudel$^{\ast}$,~\IEEEmembership{University of Rhode Island, Kingston RI.}
}

\maketitle

\begin{abstract}
Real-world biosignal data is frequently corrupted by various types of noise, such as motion artifacts, and baseline wander \cite{chatterjee2020review}. Although digital signal processing techniques exist \cite{guerrero2007biolab} to process such signals; however, heavily degraded signals cannot be recovered. In this study, we aim to classify two things: first, a binary classification of noisy and clean biosignals, and next, to categorize various kinds of noise such as motion artifacts, sensor faliue, etc.  We implemented K-means clustering, and our results indicate that the algorithm can most reliably group clean segments from noisy ones, particularly strong performance in identifying clean data compared to various categories of noise. This approach enables the selection of only high-quality bio-signal segments and provides accurate results for feature engineering that may enhance the precision of machine learning models trained on biosignals. 
\end{abstract}


%
\IEEEpeerreviewmaketitle

\section{Introduction}
%
%
%
%
\IEEEPARstart{R}{eal} world bio-signals often contain mixed and noisy data. Processing this noisy signals has always been challenging. Machine learning outcome or predictions improves  completely on the quality of input data \cite{rnmo2006electrocardiogram}. With advancements in computational models, Machine Learning technique provide a better capabilities for preprocessing such signals. The application of Machine Learning in health science posses huge potential where as it's equally challenging to execute due to poor quality input streams of bio-signals. The research in using bio-signals like ECG, PPG in predicting the cardio vascular diseases \cite{filist2024biotechnical} along with identifying withdrawal symptoms in individual with substance use disorder \cite{chapman2022impact} are one of the most trending topics to solve real world problem. Hence bio-signal processing plays a crucial role in healthcare monitoring and diagnostics.Although digital signal processing, such as FIR/IIR \cite{ahlstrom2007digital}, helps in reducing noise, there still exist gray areas or noisy segments. To aid in solving the problem that digital filter failed, we are looking into leveraging the Machine learning technique in clustering the signal into noisy or clean as our first goal. Our primary objectives includes:
\begin{itemize}
    \item To apply and observe how well clustering algorithms can identify or cluster noisy signals as noisy and clean signal as clean.
    \item To evaluate the performance of the algorithms against a ground truth labeled dataset.
\end{itemize}
In this study we developed unsupervised machine learning framework for classifying noise in bio-signals. The key contribution of this work would be comparative evaluation of multiple clustering algorithms and open-source data-set, github repository with reproducible workflow. 
\section{Related Works}
Several studies have explored various techniques for automated signal noise classifications. Rodrigues et al.\cite{rodrigues2017noise}  have implemented agglomerative clustering to segment clean and noisy ECG signal. The results were promising. Likewise, Vijaykumar et al. \cite{vijayakumar2022ecg} had implemented neural network model to classify ECG inot noisy and clean. Joseph et al. \cite{joseph2014photoplethysmogram} have implemented wavelet denoising technique to filter the additive white gaussian noise from the raw PPG signals. 
\section{Methodology}

\subsection{Physiological Data Collection Protocol}
A controlled study was designed with a standarized protocol to obtain a multimodel (ECG and PPG) dataset. Data were collected using a BIOPAC system at high frequency sampling rate of 1000 Hz. 
\begin{figure}[H]
    \centering
    \includegraphics[width=0.95\linewidth]{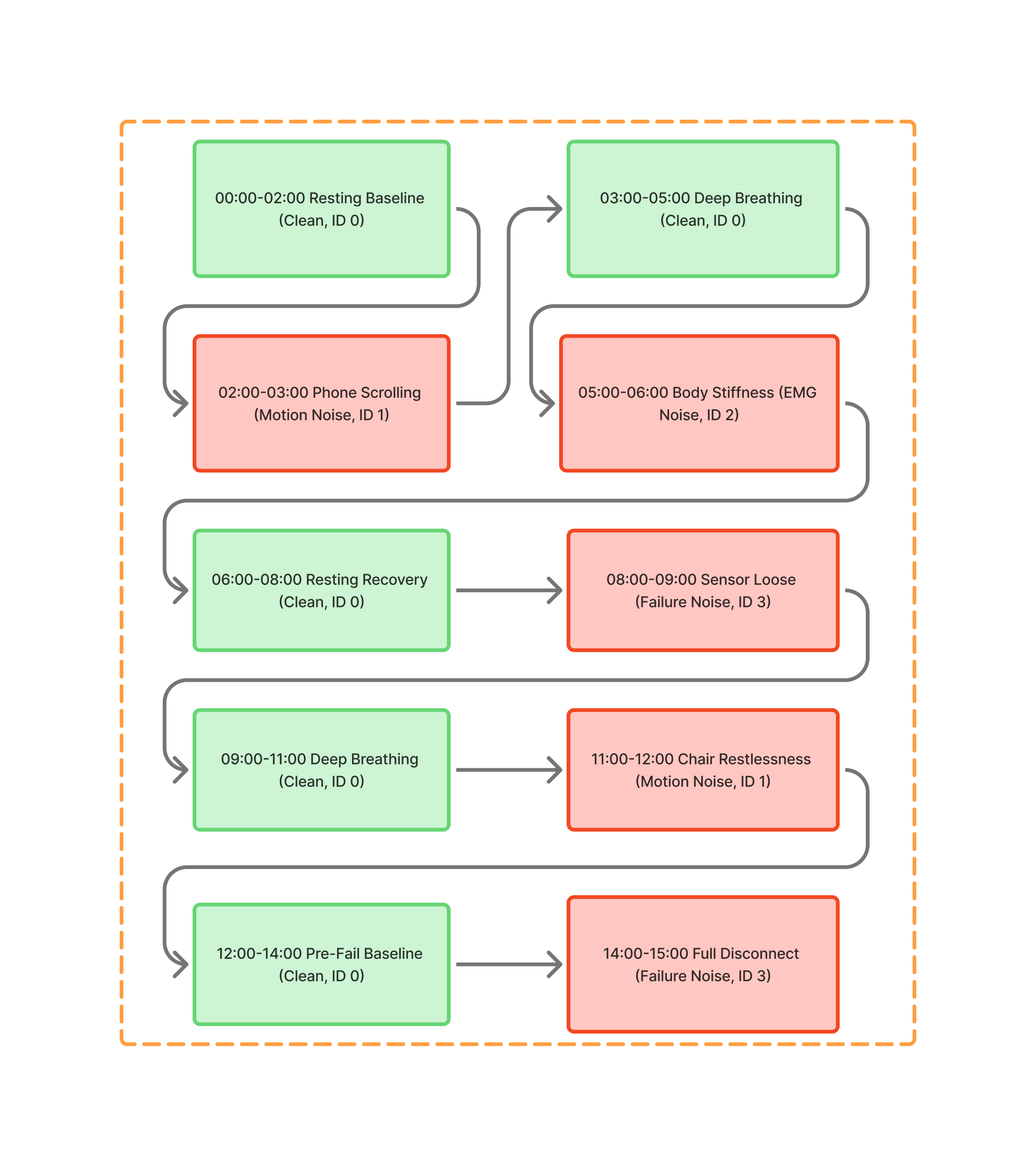}
    \caption{Physiological data collection protocol}
    \label{fig:protocol.png}
\end{figure}
We designed a specific task with standardized protocol to induce various kinds of possible noise in bio-signals. We had designed the protocol to have clean bio-signal with rest and relax and disturbed signal with sets of tasks. The activities were performed in an alternating series of 2 minutes of relaxation and 1 minute of specific activities to generate various kinds of noise. 
All artifact classes were assigned discrete integer labels for downstream evaluation of clustering and classification models. The ground truth label was assigned, ``Label~0'' represented physiologically valid segments (resting and deep breathing), ``Label~1'' denoted motion-related activities, ``Label~2'' corresponded to high-frequency, muscles or EMG noise, and ``Label~3'' marks the sensor failure or disconnection events. This structured dataset provides a controlled benchmark for assessing the ability of unsupervised algorithms to differentiate true physiological dynamics from heterogeneous noise sources.
\begin{figure}[H]
    \centering
    \includegraphics[width=0.9\linewidth]{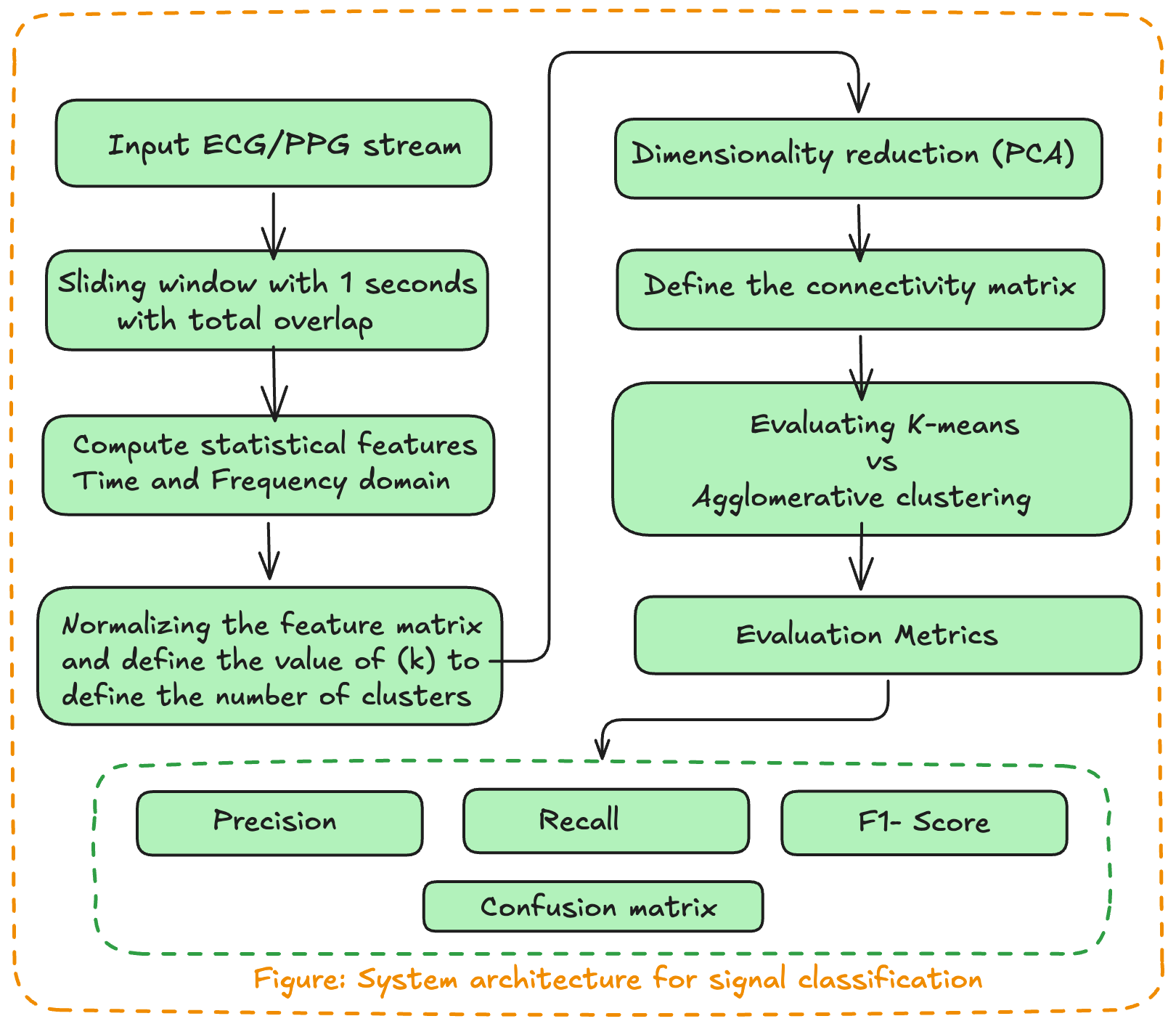}
    \caption{Proposed System Flow}
    \label{fig:system_arch}
\end{figure}

We implemented the standard machine learning method, which includes a standard procedure from data preprocessing to model evaluations. 
The different tools and packages that were implemented includes. 
1.	Pandas: For data manipulations. 
2.	Matplotlib and seaborn: For graph creation or data visualizations
3.	Scikit-learn: for data preprocessing, modeling and evaluations.

The following process workflow helps to explain the method more descriptively. 
\subsection{Data preprocessing}
	The dataset would be checked if it contains any null values or inconsistencies in data format. Any null values or discontinued time series data would be discarded. 
    
\subsection{Model selection:}
We classified signals into different clusters using two different clustering methods, to check their performance. The agglomerative clustering techniques were taken reference from \cite{rodrigues2017noise} authors. We evaluated the signal separation from K mean and compared it against the hierarchical clustering to evaluate the comparative performance. 
\begin{enumerate}
    \item K-Means Clustering
    \item Agglomerative Clustering
\end{enumerate}

\subsection{Evaluation Metrics} 
To compute overall performance, we will computed comprehensive metrics that include:
\begin{enumerate}
    \item \textbf{Accuracy:} Measures how often a model makes a correct prediction out of all predictions.
    \item \textbf{Precision:} Measures the accuracy of positive predictions.
    \item \textbf{Recall:} Measures the model's ability to find all positive instances.
    \item \textbf{F1-Score:} The harmonic mean of precision and recall.
    \item \textbf{Confusion Matrix:} A visualization table that evaluates the performance of a classification model, reflecting the correct and incorrect predictions for each class.
\end{enumerate}

\subsection{Data description}
The physiological data is collected from multiple sources. 
\begin{enumerate}
    \item Open source database, physionet \cite{moody2022physionet}
    \item Data collected from wearable and biosensing lab (WBL), University of Rhode island. 
\end{enumerate}
We have collected the physiological data from wearable biosensing lab, at Univeristy of rhode island. The collected data contains (ECG, and PPG) data along with the ground truth on labels for noise and clean segments. We had set a protocol of 15 minutes with several activites to induce various form of noise and label those ground truth to evaluate model performance.Apart from that there are publicly available dataset available on \cite{moody2022physionet} physionet database which can be used for n iterations. On our preliminary experiement we have only evaluated the data collected from WBL lab.

\subsection{Feature extractions}
We implemented both time and frequency domain features to let model understand the patterns in signals in more identifiable signature for better results. We implemented  a 2-minute sliding window to extract features and used silhouette score to identify which value of K for clustering provides a better results.
\begin{figure}[H]
    \centering
    \includegraphics[width=0.9\linewidth]{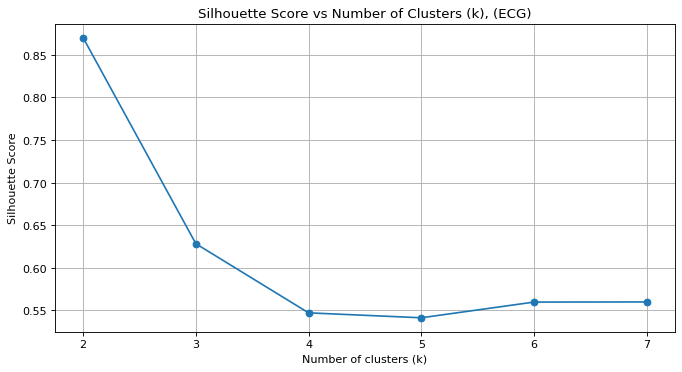}
    \caption{Silhouette score across different values of k for identifying the optimal cluster}
    \label{fig:Silhouette score}
\end{figure}
We experiment the value k = 4 for multi class noise separation.
The table below provides details on the lists of  time-domain and frequency-domain features we implemented to feature engineer for our model \cite{singh2023ecg}.

\begin{table}[!htbp]
\centering
\caption{Statistical Features for ECG Morphology Clustering}
\label{tab:morpho_features}
\renewcommand{\arraystretch}{1.3} 
\begin{tabular}{lp{0.65\columnwidth}} 
\toprule
\textbf{Feature} & \textbf{Description} \\
\midrule

\multicolumn{2}{l}{\textit{\textbf{Time-Domain Statistics}}} \\
\addlinespace[3pt] 
\multicolumn{2}{l}{\textit{Amplitude Statistics}} \\
\quad Mean & Average amplitude value. \\
\quad Variance & Measures the spread of amplitude values. \\
\quad Median & The 50th percentile of amplitude values. \\
\addlinespace[3pt]
\multicolumn{2}{l}{\textit{Shape Statistics}} \\
\quad Skewness & Measures asymmetry. \\
\quad Kurtosis & Measures peakiness. \\
\addlinespace[3pt]
\multicolumn{2}{l}{\textit{Noise and Activity Statistics}} \\
\quad ZCR & \textbf{Zero-Crossing Rate:} Number of times signal crosses zero. \\
\quad Energy & \textbf{Signal Energy:} Sum of squared amplitude values; measures overall power in the window. \\
\midrule

\multicolumn{2}{l}{\textit{\textbf{Frequency-Domain (Spectral) Statistics}}} \\
\addlinespace[3pt]
\multicolumn{2}{l}{\textit{Spectral Power Statistics}} \\
\quad Total Power & Sum of power across all frequencies in the Power Spectral Density (PSD). \\
\quad Power (>30 Hz) & Power in the high-frequency noise band; high value indicates Muscle Artifacts (EMG). \\
\addlinespace[3pt]
\bottomrule
\end{tabular}
\end{table}

\subsection{PCA: dimensionality reduction}
Apart from clustering the ECG waveforms, we have also computed the dimensionality reduction for currently implemented 9 different features into lower dimension. We computed  time-domain features per ECG window (mean, variance, median, skewness, kurtosis and zero crossing rate, root mean square ) and frequency domain features including (power spectral density and signal power at frequency band higher than 30 Hz). Evaluating the significance of features into 9 dimension was reduced into 2 dimension using Principal component analysis (PCA) which find the direction of maximum variance in the data. 
\begin{figure}[H]
    \includegraphics[width=0.9\linewidth]
    {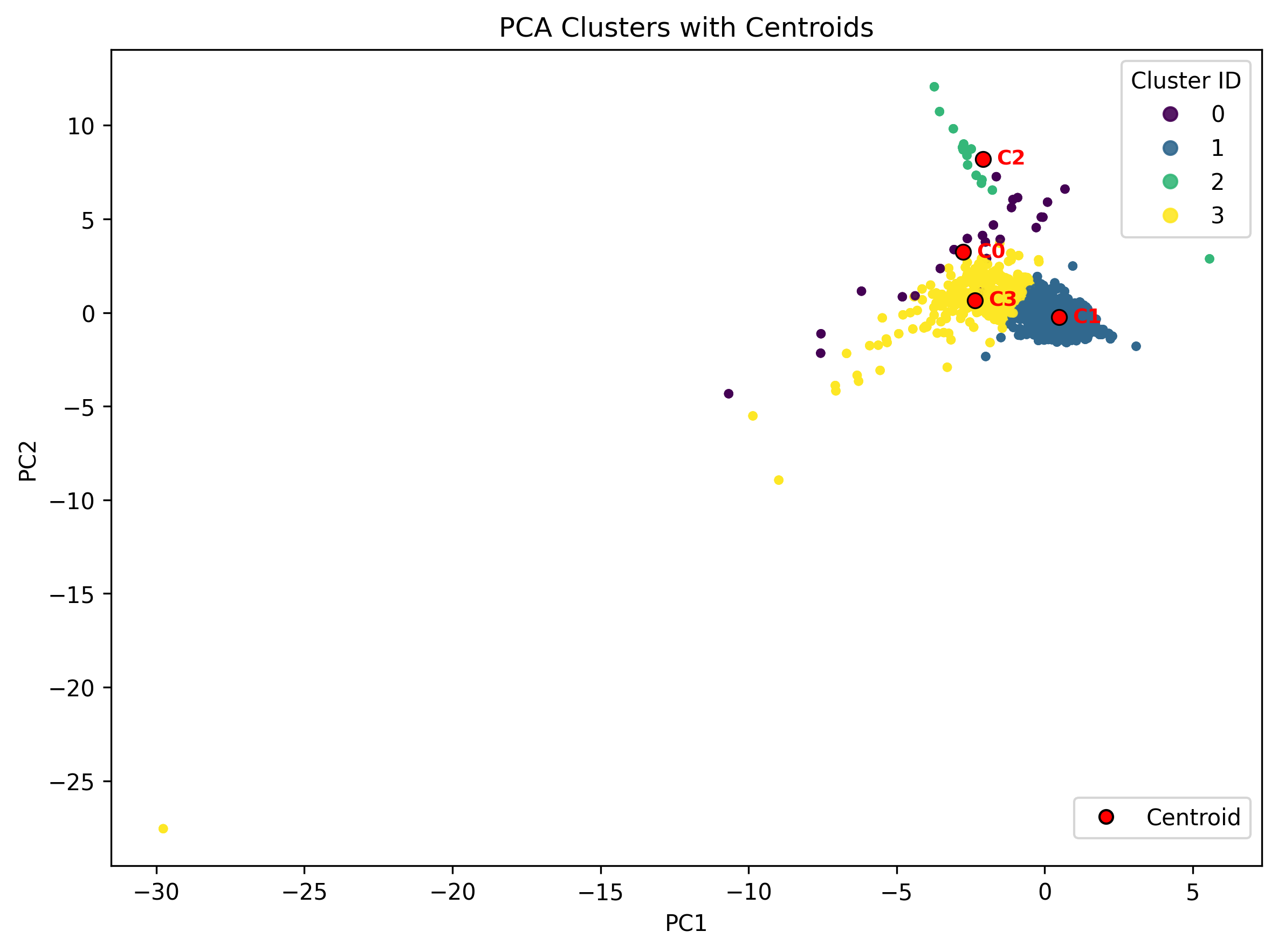}
    \caption{PCA: Dimensionality reduction using PCA}
    \label{fig:K-mean clustering with centroids}
\end{figure}
We obtained a variance ratio for both PC1 and PC2 to be [0.55,0.22]. PC1 explains 55 percentage of the total variance while PC2 explains 25 percentage of it. Total 80 percentage of 9-D features space is visible into this 2-D plot. 


\section{Results and Discussion}
\subsection{Multi-class noise type clustering (ECG) }
\begin{figure}[H]
    \centering
    \includegraphics[width=0.9\linewidth]{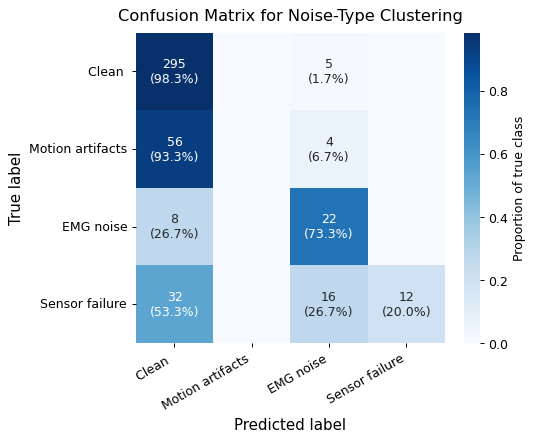}
    \caption{Confusion matrix for ECG noise-type clustering.}
    \label{fig:ecg_confusion_multiclass}
\end{figure}
The multi-class confusion matrix in Fig \ref{fig:ecg_confusion_multiclass} summarizes how purposed K-mean framework helps in separating different ECG noise types. The clean signal class was most accurately separated (98.3\%). However, the motion artifacts were unable to be classified. This is because the electrode were placed on the chest region which are relatively resistant to action of light motion like scrolling cell-phone. Although the activities were labeled for motion it did not significantly changed the waveform. This finding  suggests us that smaller movements doesn't affects heavily in signal morphology. The model performed moderate result in classifying the EMG or muscles noise with (73.3\%) correct classification. The sensor failure was also weekly classified. The performance of noise classification can be improved by increasing the number of k value. 
\begin{figure}[H]
    \centering
    \includegraphics[width=0.95\linewidth]{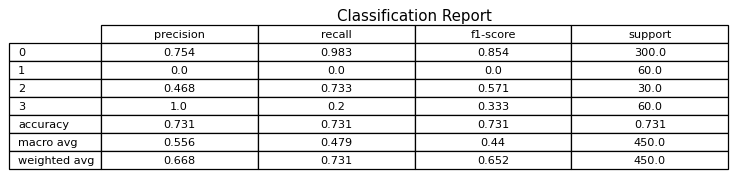}
    \caption{Per-class precision, recall, and F1-score for ECG clustering.}
    \label{tab:ecg_classification_report}
\end{figure}
Table ~ \ref{tab:ecg_classification_report} reports the precision, recall, and F-1 score for each type of noise. The clean class (Class 0) attains high precision (0.754) and very strong recall (0.983), confirming that clean ECG segments are almost always recognized correctly. In contrast, motion artifacts (Class 1)  obtained near-zero precision and recall, meaning the current feature set and clustering approach fail to capture their subtle distortions  created by motion. EMG noise (Class 2) achieves moderate precision (0.468) and recall
(0.733), consistent with the clearer spectral signature of muscular
activity. Sensor failure (Class 3) presents high precision (1.0) but low recall
(0.20); whenever the algorithm predicts sensor failure it is almost
always correct, but it misses the majority of such windows.

The macro-averaged F1-score (0.44) highlights the imbalance in
performance across classes, while the weighted F1-score (0.652) is
dominated by the large clean class. These results emphasize that
although the model is effective at preserving clean data, additional
discriminative features (e.g., band-limited power ratios, baseline
wander metrics, or shape descriptors) are required to reliably separate
all noise categories, particularly motion artifacts and subtle sensor
instability. Apart from that the number of cluster should be increased upto 6 to have better gain in clustering noisy segments.
\begin{figure}[H]
    \centering
    \includegraphics[width=0.7\linewidth]{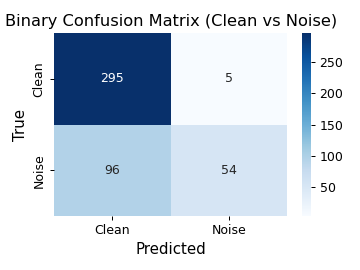}
    \caption{Binary confusion matrix for clean vs.\ noise ECG classification.}
    \label{fig:ecg_confusion_binary}
\end{figure}
We also evaluated the clustering framework into a system to evaluate noisy vs clean windows. We merged all the noisy category to evaluate the practical usefulness of the clustering algorithm. The resulting binary confusion matrix in Fig. !\ref{fig:ecg_confusion_binary} shows a true reflection of system performance in segmenting clean and noisy segment. The algorithm correctly classifies 295 windows as clean and incorrectly flags the 5 clean windows as noise, which reflects algorithm to preserve the clean and valid data. 
On the noisy side, 54 windows were correctly identifed as noise while 96 windows are misclassified as clean. The system, was not great in discarding the waveform with mild artifacts. 
\subsection{Multi-class noise type clustering (PPG) }
\begin{figure}[H]
    \centering
    \includegraphics[width=0.75\linewidth]{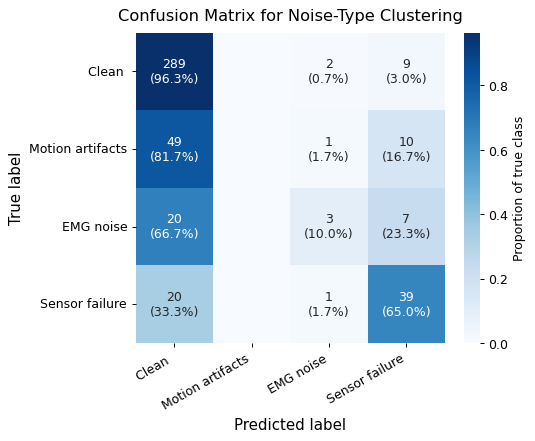}
    \caption{Confusion matrix for PPG noise-type clustering.}
    \label{fig:ppg_confusion_multiclass}
\end{figure}
The multi-class confusion matrix in Fig.~\ref{fig:ppg_confusion_multiclass} presents how the K-means algorithm separates different types of cotegories of PPG signal. Similar to ECG, the clean PPG class is identified with high reliability (96.3\%),representing the clustering algorithm to capture physiologically valid waveforms with higher accuracy. However, motion artifacts were not well distinguished, with 81.7\% of motion-affected windows misclassified as clean. This is expected, as mild finger movements often produce subtle distortions that preserve the overall pulse shape, making them difficult to detect through statistical features alone.

EMG-related muscle noise poor separation, with 10.0\% correctly classified. Muscle tension typically introduces high-frequency components and baseline fluctuations in PPG, but these signatures are weaker and more inconsistent than in ECG, contributing to the misclassification of the remaining windows. Sensor failure (Class~3) achieved improved detection compared to ECG, with 65\% correctly identified. PPG sensors tend to exhibit abrupt amplitude collapse or sudden waveform distortion during detachment, which produces more separable feature patterns in the clustering space.

\subsection{Per-Class Precision, Recall, and F1-Score for PPG}

\begin{figure}[H]
    \centering
    \includegraphics[width=0.9\linewidth]{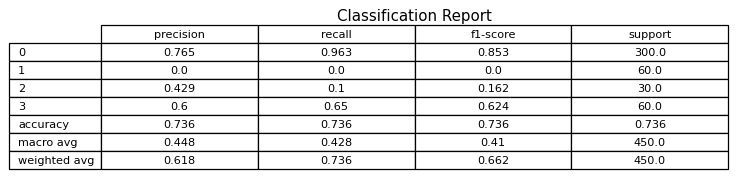}
    \caption{Confusion matrix for PPG noise-type clustering.}
    \label{fig:ppg_table_report}
\end{figure}
The per-class evaluation in Table~\ref{fig:ppg_table_report} further highlights the strengths and weaknesses of the PPG clustering pipeline. Clean PPG windows (Class~0) achieve high precision (0.765) and recall (0.963), confirming that the algorithm reliably preserves valid physiological data. However, motion artifacts (Class~1) again exhibit near-zero precision and recall, consistent with their subtle and diverse distortion patterns that are not captured effectively by the current feature set.

EMG noise (Class~2) shows low recall (0.10), indicating that most muscle-induced disturbances remain unrecognized. These distortions generate high-frequency spectral content, but their manifestation in PPG is variable and often weak, making them difficult to isolate without more specialized spectral descriptors. Sensor failure (Class~3) demonstrates moderate detection performance, with precision of 0.60 and recall of 0.65, supported by the clearer morphological disruptions that occur during partial or complete sensor detachment.

The macro-averaged F1-score (0.41) reflects the imbalance across noise types, while the weighted F1-score (0.662) is heavily influenced by the large clean class. Overall, these results suggest that although the clustering model performs strongly on clean data, additional checks with different k values and features should be tested for better results.

\subsection{Binary Clean vs.\ Noise Classification (PPG)}
\begin{figure}[H]
    \centering
    \includegraphics[width=0.75\linewidth]{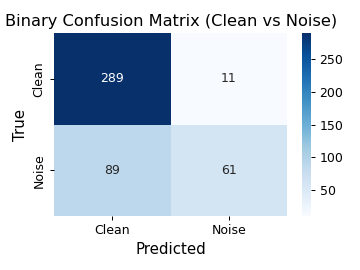}
    \caption{Confusion matrix for PPG binary clustering.}
    \label{fig:ppg_binary_clustering}
\end{figure}
To assess practical usefulness, for PPG all noise categories were merged into a single ``noisy'' class. The binary confusion matrix (Fig.~\ref{fig:ppg_binary_clustering}) shows that the clustering method performs substantially better when used as a clean-versus-noisy detector rather than a multi-class classifier. The framework correctly identifies 289 out of 300 clean windows, misclassifying only 11 as noise, highlighting its strong ability to retain physiologically valid PPG segments.

For noisy segments, the system correctly identifies 61 windows while misclassifying 89 as clean. This indicates that the method is effective at preserving clean PPG but struggles to reject mild artifact segments particularly those associated with small finger movements. Compared to ECG, PPG displays similar overall accuracy but increased difficulty in distinguishing mild motion artifacts, reflecting the inherently motion-sensitive nature of optical PPG sensing.

Despite these challenges, the binary evaluation confirms that the unsupervised clustering framework can still provide meaningful segmentation of clean versus noisy PPG windows, making it a valuable preprocessing step for downstream physiological modeling or machine learning tasks.

Hence, our current implemented methodology works superior in segmenting bio- signals into clean and noisy segments, however struggles in accurately classifying the various kinds of noise. Current method helps in segmenting morphological valid and clean signal which can results in more accurate results when the signal is used in any Machine learning architecture in performing desired analysis.

\section{Project source code and repository}
The source code and dataset is publicly available via our GitHub repository:\url{ https://github.com/paudel54/Clustering_ML}


\section*{Acknowledgment}
I would like to express sincere gratitude to Professor
Alina Barnett of the University of Rhode Island for her valuable guidance guidance, constructive feedback, and continuous support throughout the development of this work.

\appendix 
\section{ECG signal separation using clustering }
This appendix presents some results  of ECG and PPG windows clustered into different identifed clusters. These example provides insights on how powerful the clustering algorithm be in segmenting clean segments from noisy one. 

\begin{figure}[H]
    \centering
    \includegraphics[width=0.9\linewidth]{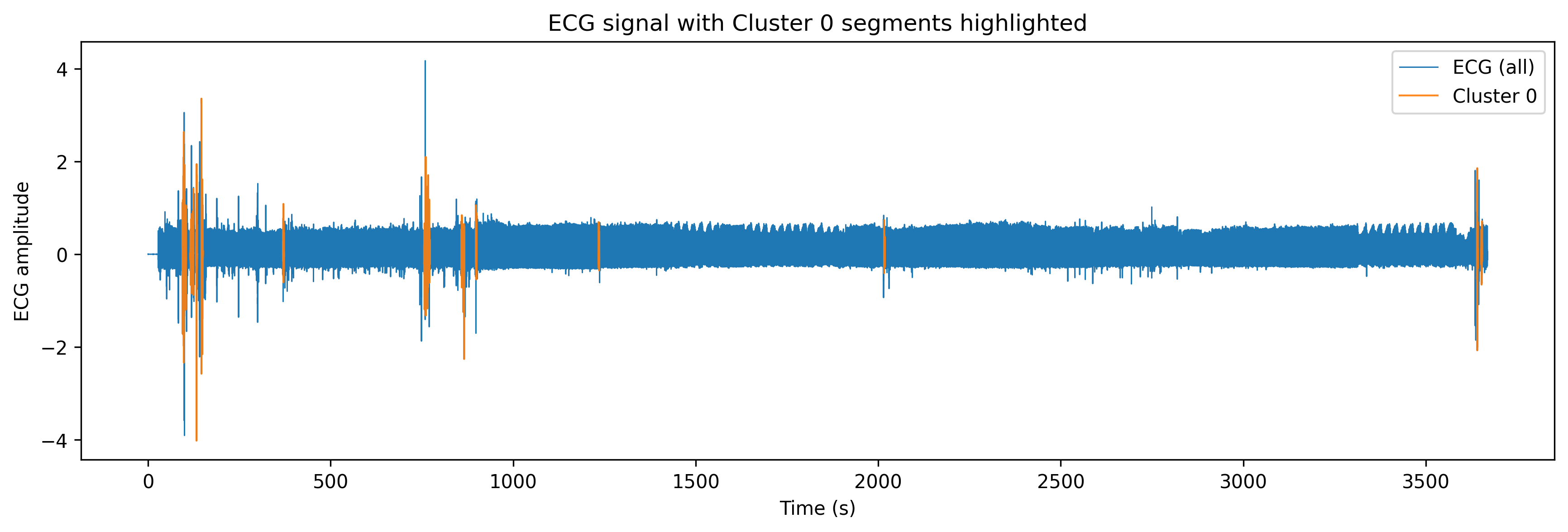}
    \caption{ECG: Cluster 0 waveform representing motion artifacts}
    \label{fig:ecg_cluster_0}
\end{figure}

\begin{figure}[H]
    \centering
    \includegraphics[width=0.9\linewidth]{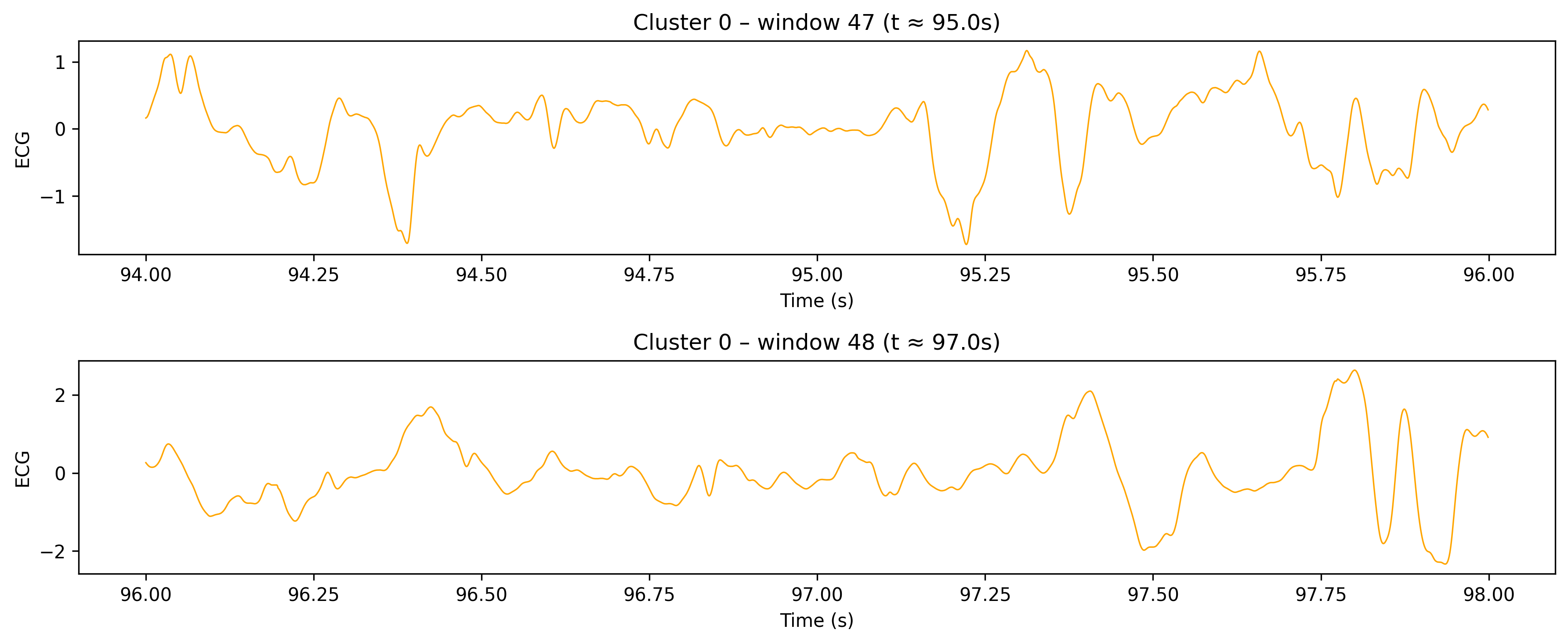}
    \caption{ECG: Cluster 0 sample window (motion artifacts )}
    \label{fig:ecg_window_0}
\end{figure}

\begin{figure}[H]
    \centering
    \includegraphics[width=0.9\linewidth]{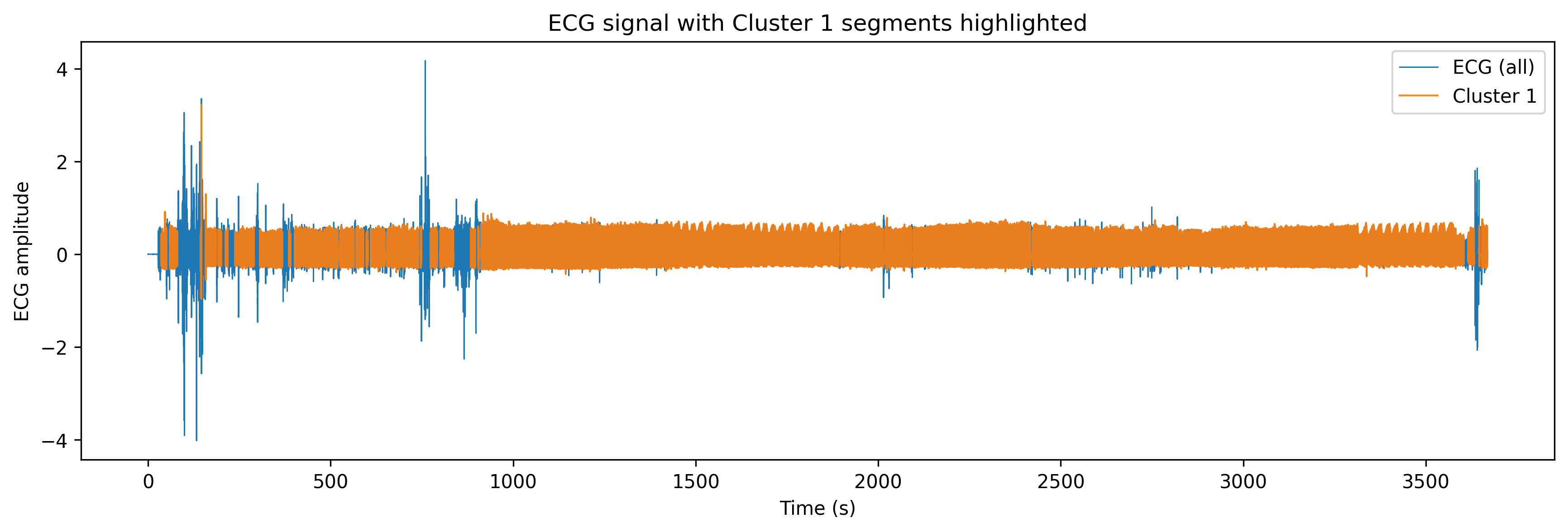}
    \caption{ECG: Cluster 1 waveform representing stable and clean signal}
    \label{fig:ecg_cluster_1}
\end{figure}

\begin{figure}[H]
    \centering
    \includegraphics[width=0.9\linewidth]{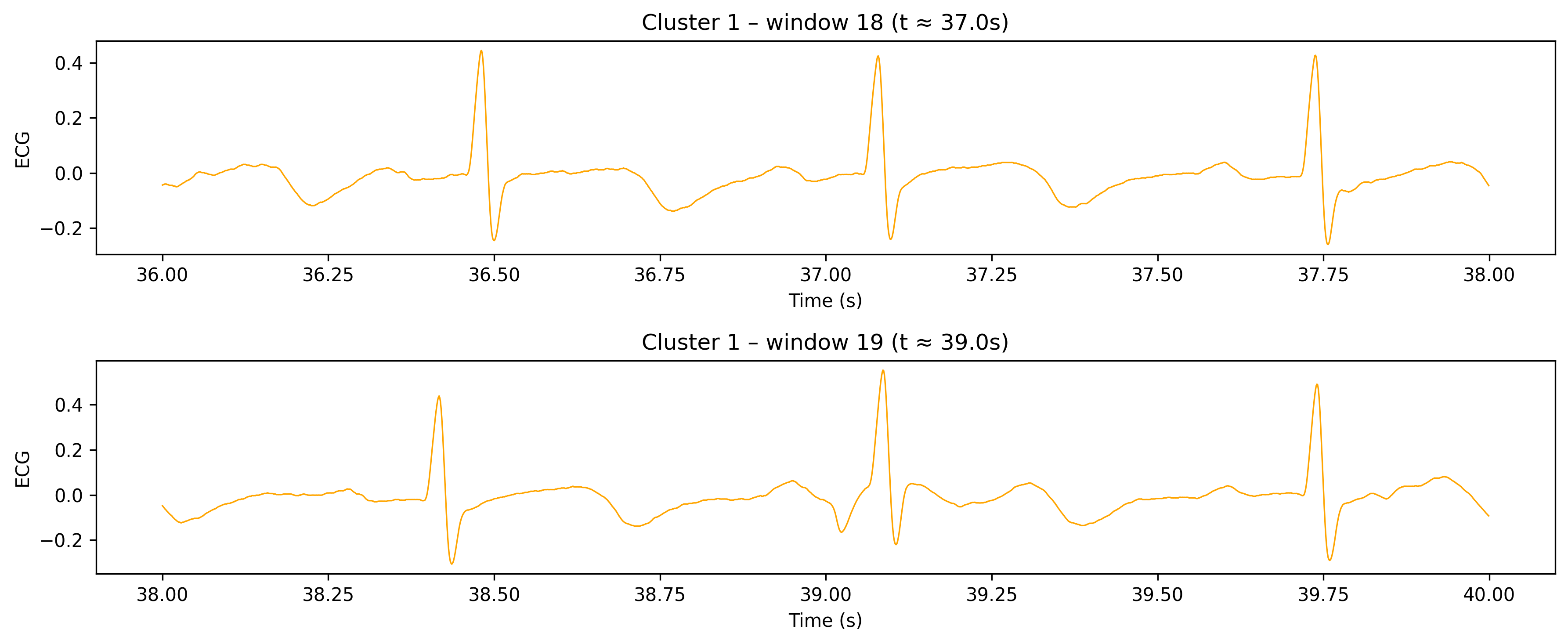}
    \caption{ECG: Cluster 1 sample window (clean waveform)}
    \label{fig:ecg_window_1}
\end{figure}

\begin{figure}[H]
    \centering
    \includegraphics[width=0.9\linewidth]{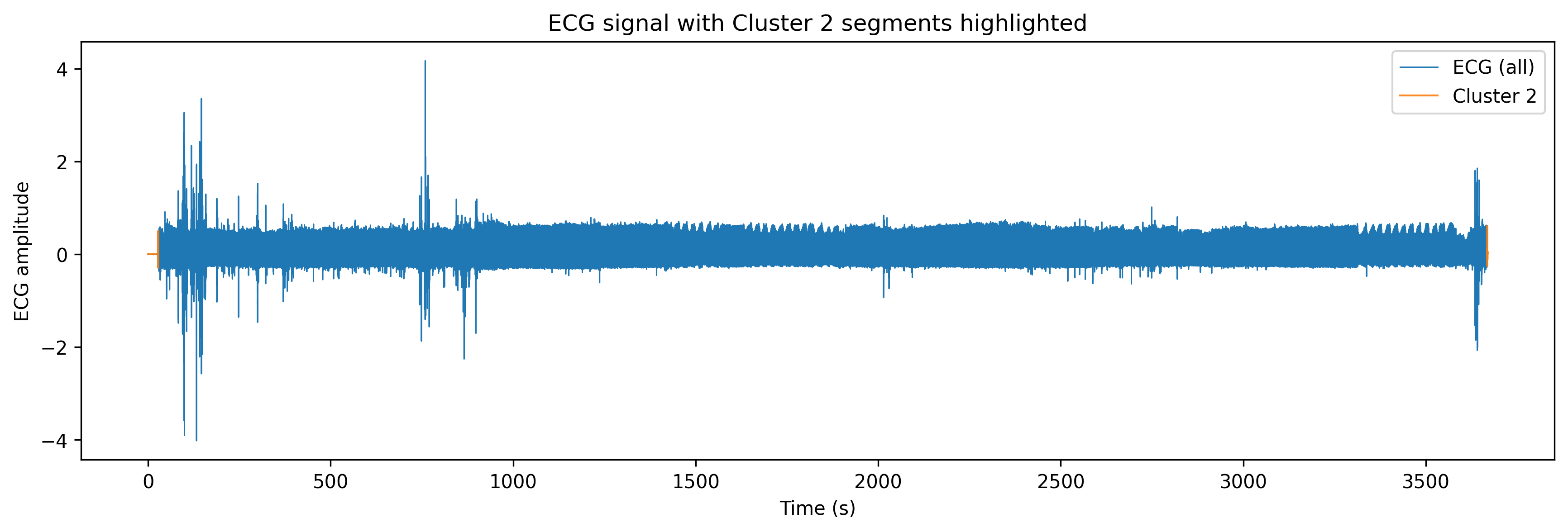}
    \caption{ECG: Cluster 2 waveform represeting sensor noise}
    \label{fig:ecg_cluster_2}
\end{figure}

\begin{figure}[H]
    \centering
    \includegraphics[width=0.9\linewidth]{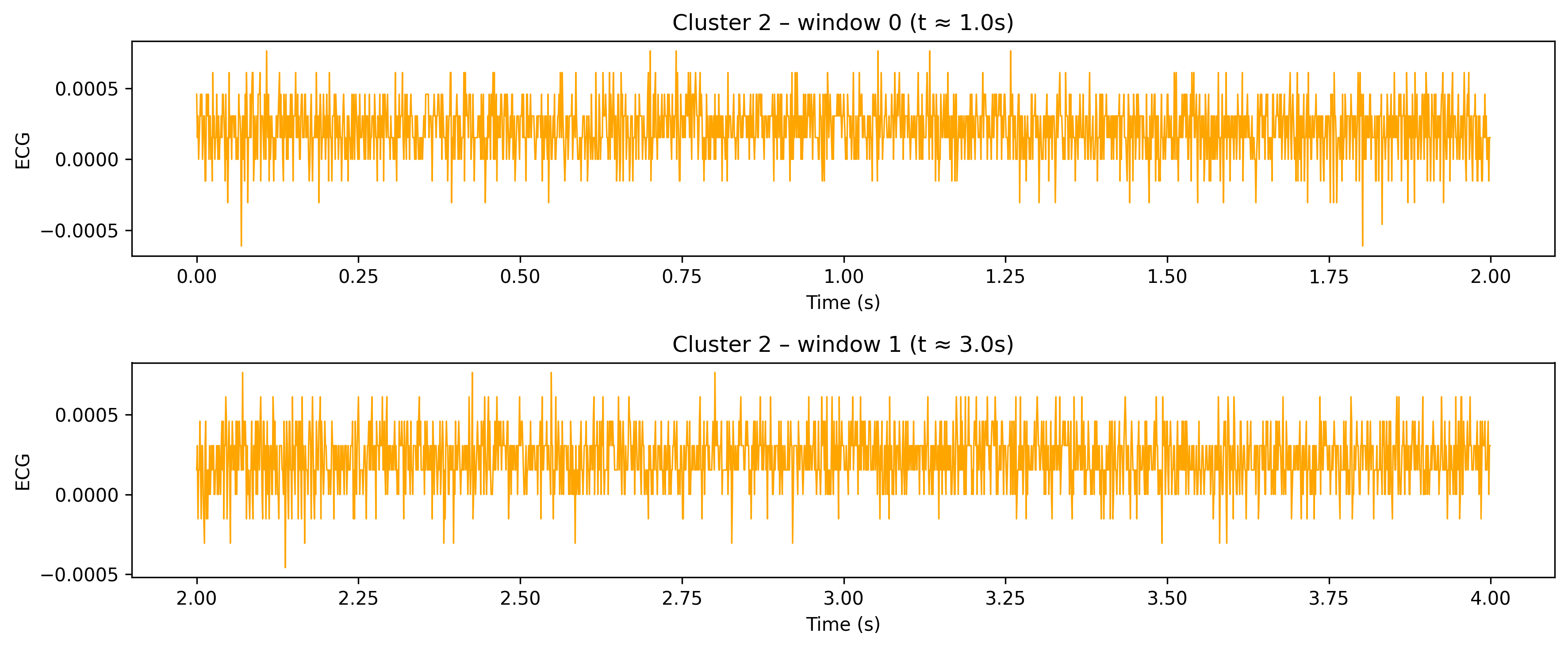}
    \caption{ECG: Cluster 2 sample window (sensor noise)}
    \label{fig:ecg_window_2}
\end{figure}




\begin{figure}[H]
    \centering
    \includegraphics[width=0.9\linewidth]{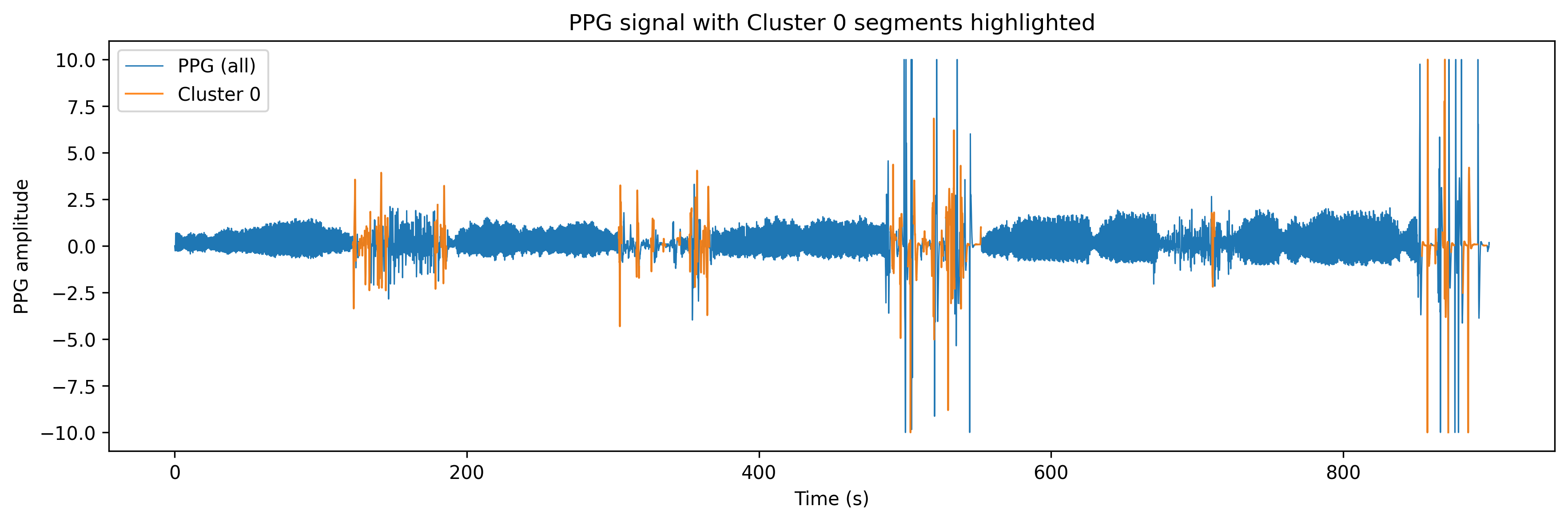}
    \caption{PPG: Cluster 0 representing EMG noise }
    \label{fig:ppg_cluster_0}
\end{figure}

\begin{figure}[H]
    \centering
    \includegraphics[width=0.9\linewidth]{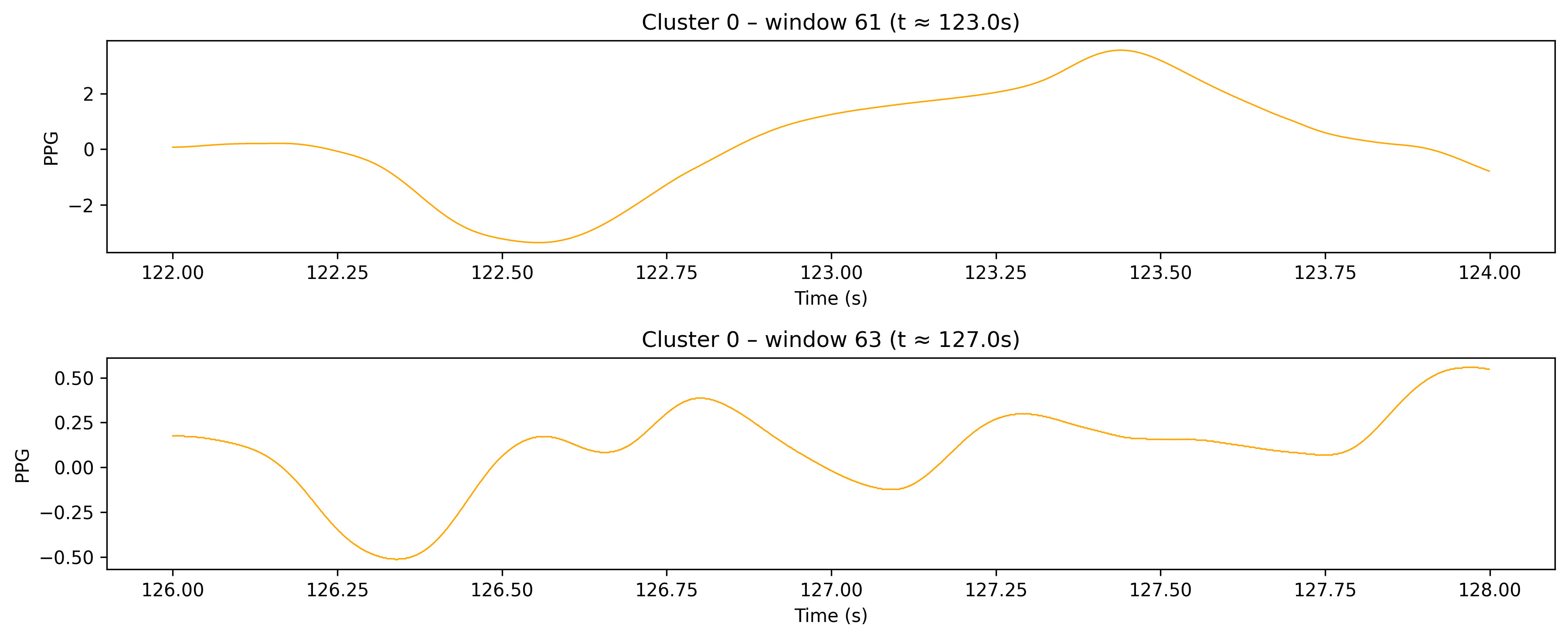}
    \caption{PPG: Cluster 0 with (EMG waveform)}
    \label{fig:ppg_window_0}
\end{figure}

\begin{figure}[H]
    \centering
    \includegraphics[width=0.9\linewidth]{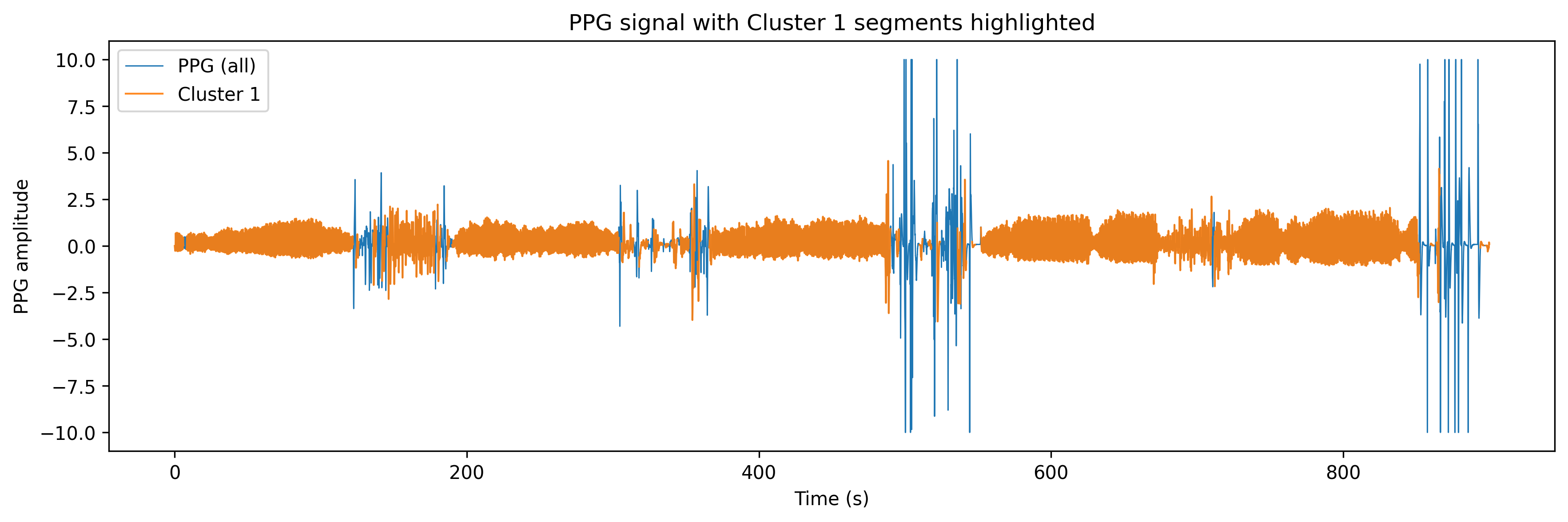}
    \caption{PPG: Cluster 1 (clean PPG) }
    \label{fig:ppg_cluster_1}
\end{figure}

\begin{figure}[H]
    \centering
    \includegraphics[width=0.9\linewidth]{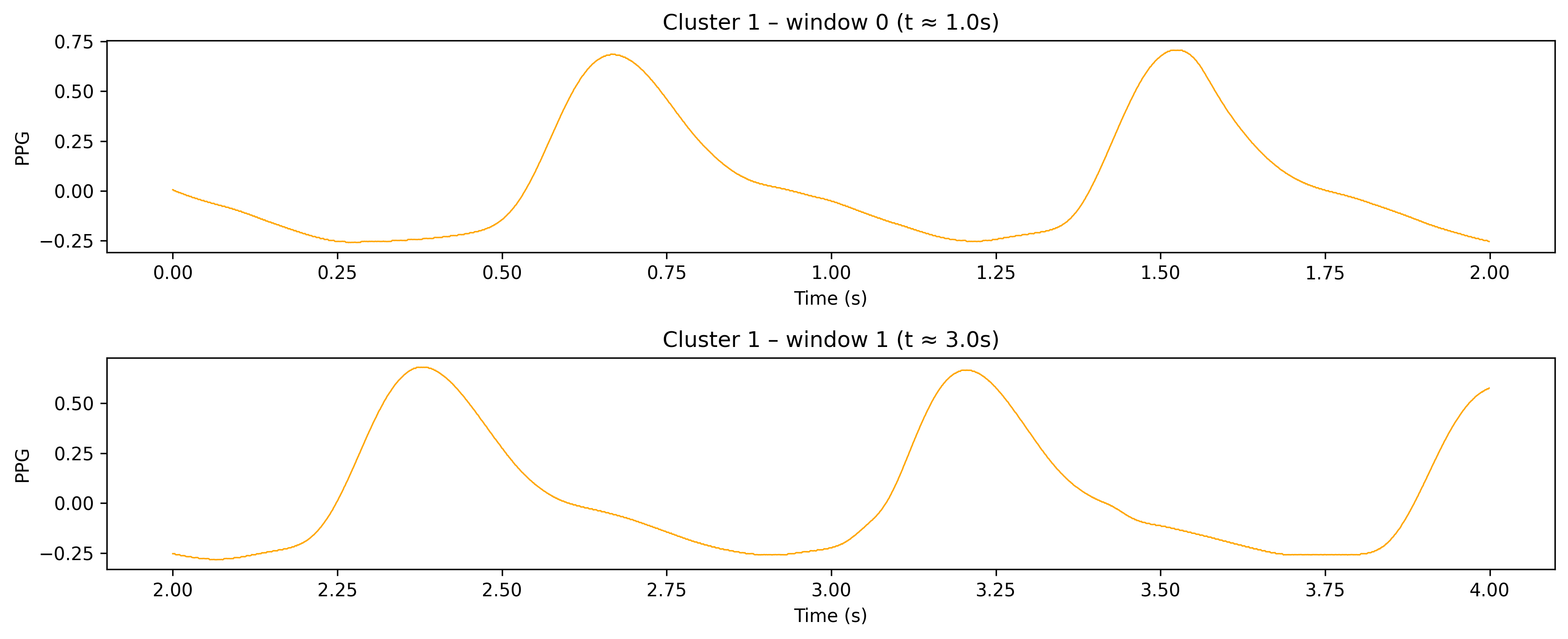}
    \caption{PPG: Cluster 1 representing clean PPG windows}
    \label{fig:ppg_window_1}
\end{figure}

\begin{figure}[H]
    \centering
    \includegraphics[width=0.9\linewidth]{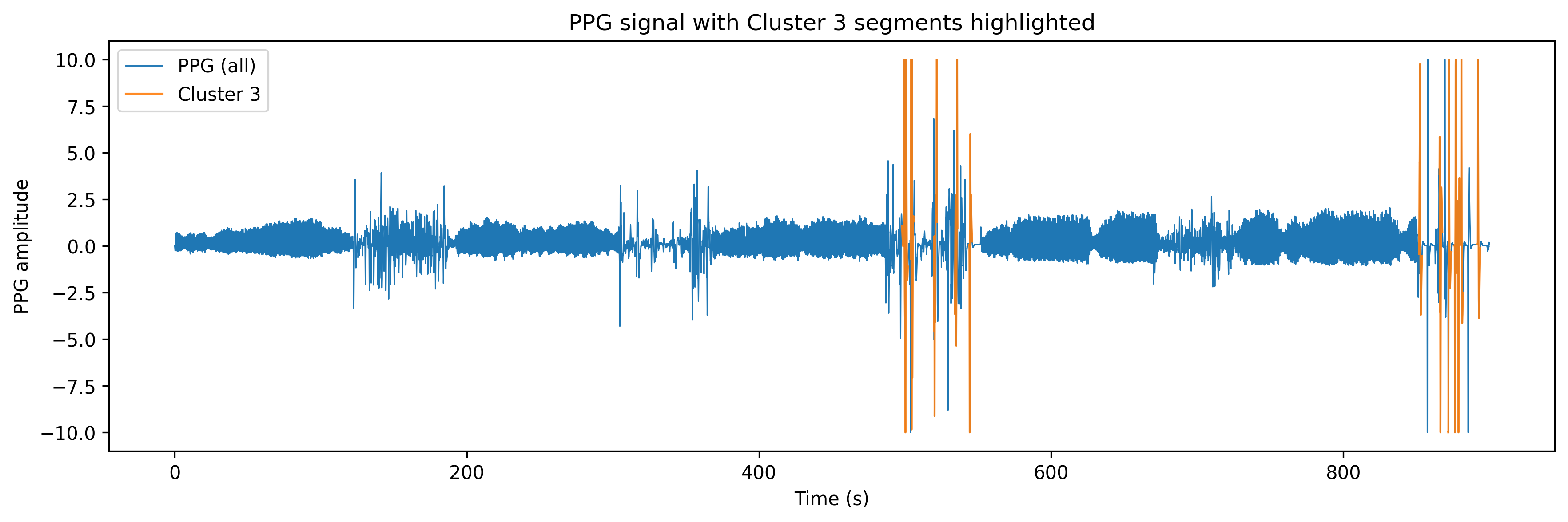}
    \caption{PPG: Cluster 3 representing motion artifacts}
    \label{fig:ppg_cluster_3}
\end{figure}

\begin{figure}[htbp]
    \centering
    \includegraphics[width=0.9\linewidth]{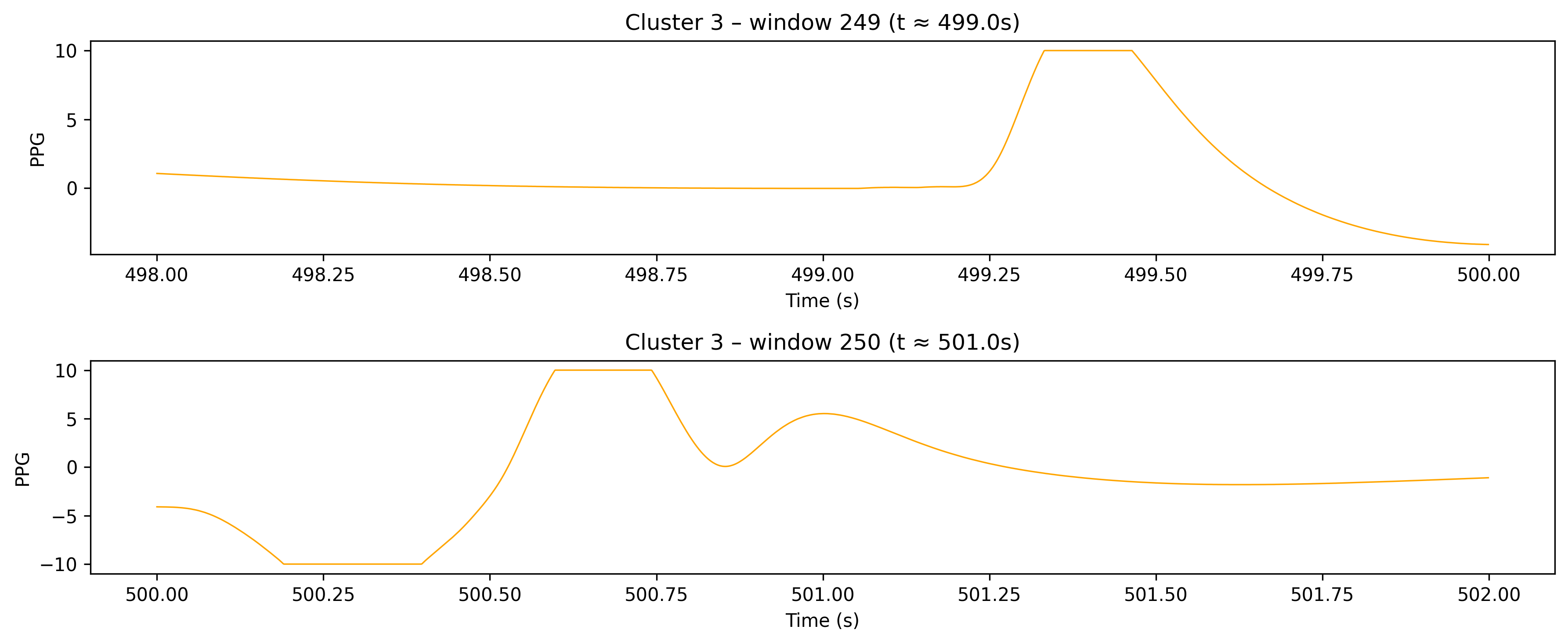}
    \caption{PPG: Cluster 3 ( motion artifacts) }
    \label{fig:ppg_window_3}
\end{figure}

\bibliographystyle{IEEEtran}
\bibliography{references}

\end{document}